\documentclass[twocolumn,showpacs,amsmath,amssymb,pre]{revtex4}

\usepackage{amsmath,amssymb,latexsym,graphics,epsfig}
\usepackage{pstricks}
\usepackage{float}

\newpsobject{showgrid}{psgrid}{subgriddiv=1}

\newcommand \be {\begin{equation}}
\newcommand \ee {\end{equation}}
\newcommand \bea {\begin{eqnarray}}
\newcommand \eea {\end{eqnarray}}

\begin{document}

\title{Order statistics of $1/f^\alpha$ signals}

\author{N. R. Moloney}
\email{moloney@pks.mpg.de}

\affiliation{Max Planck Institute for
the Physics of Complex Systems, N\"othnitzer Str. 38, 
Dresden, D-01187 Germany}

\author{K. Ozog\'any}
\email{ozogany@general.elte.hu}

\author{Z. R\'acz}
\email{racz@general.elte.hu}
\affiliation{Institute for Theoretical Physics - HAS,
  E\"otv\"os University, P\'azm\'any
  s\'et\'any 1/a, 1117 Budapest, Hungary}

\date{\today}

\begin{abstract}
Order statistics of periodic, Gaussian noise with $1/f^\alpha$ power
spectrum is investigated. Using simulations and phenomenological 
arguments, we find three scaling regimes for
the average gap $d_k=\langle x_k-x_{k+1}\rangle$ 
between the $k^{th}$ and $(k+1)^{st}$ largest values of the 
signal. The result $d_k\sim k^{-1}$ known for independent, 
identically distributed variables remains valid for $0\le \alpha <1$.
Nontrivial, $\alpha$-dependent scaling exponents 
$d_k\sim k^{(\alpha -3)/2}$ emerge 
for $1 < \alpha <5$ and, finally, $\alpha$-independent scaling,
$d_k\sim k$ is obtained for $\alpha>5$. 
The spectra of average ordered values 
$\varepsilon_k=\langle x_1-x_k \rangle\sim k^\beta$ is also examined. 
The exponent $\beta$ is derived from the gap scaling as well as 
by relating $\varepsilon_k$ 
to the density of near extreme states. Known results for 
the density of near extreme states combined with scaling 
suggest that $\beta(\alpha=2)=1/2$, 
$\beta(4)=3/2$, and $\beta(\infty)=2$ are exact values. We also show that  
parallels can be drawn between $\varepsilon_k$ and the quantum mechanical
spectra of a particle in power-law potentials. 
\end{abstract}

\maketitle

\section{Introduction}
Extreme value statistics (EVS) was first developed in mathematics
\cite{FisherTippett,Galambos,deHaan}. Its importance was soon
recognized and emphasized in engineering \cite{Gumbel,Weibull},
followed by finance and environmental problems
\cite{Embrecht,Katz2002,Storch2002,Bunde2006,Gutenberg1944}.  Although
applications in physics appeared relatively late, they cover a wide
range of fields including cosmology \cite{Peebles,Lin2010}, spin
glasses \cite{Mezard}, random fragmentation \cite{Majum-random-frag},
percolation \cite{Bazant}, random matrices \cite{Majum-RM}, and, most
actively studied at present, interface fluctuations
\cite{Raychaudhuri,Gyorgyi2003,Majum2004,Guclu,Rosso2004,Leed2EW,MajumFSSSchehr,Gyorgyi2007}.

The extreme value in a batch of data is important, but its study makes
use of only a small fraction of the available information.
Accordingly, there have been various attempts to extend studies
towards near extreme characteristics, such as density of states near
extremes \cite{Sabha-Majum,Burkhardt}, first-passage and return-time
statistics \cite{Redner1st,Leadbetter}, persistence \cite{Majpersist},
and record statistics \cite{Redner-rec,Krug-rec1,Krug-rec2}. A natural
extension (which will be the concern in this paper) is to consider not
only the extreme, but the sequence $x_1,x_2,...,x_k,...$ of the
$1^{st}$, $2^{nd}$, ..., $k^{th}$,... largest, i.e. extract information
from the order statistics of the system.

Order statistics has been much studied in mathematics
\cite{Leadbetter,Pickands}. All relevant quantities are known for
independent, identically distributed (i.i.d.) variables, and a
significant amount is also known for weakly correlated systems
\cite{Leadbetter}. In physics, meanwhile, results related to order
statistics are scarce: the order statistics of the brightest galaxies
was recently proposed to replace the use of standard candles
\cite{Csabai2011}.  Another cosmological example \cite{Tremain1977}
concerns an inequality satisfied by i.i.d. variables for the ratio of
the average gap $\langle x_1-x_2 \rangle$ and the standard deviation
$\sigma_1$ of $x_1$.  Violation of this inequality by the brightness
data of galaxies leads to the notion that the brightest galaxies in
clusters are special, in the sense that their brightness cannot simply
be statistically attributed to the tail end of the luminosity
distribution.  An interesting example of order statistics in
statistical physics concerns the positions of the $k$ rightmost points
of branching random walks \cite{Derrida-branching}.  The importance of
this work is that it provides the exact order statistics of a system
of strongly correlated particles.

For applications, it is clear that a better understanding of near
extreme properties in correlated system is required. Here, we shall
make steps in this direction by investigating the order statistics of
Gaussian signals $x(t+T)=x(t)$ of period $T$ with $1/f^\alpha$ power
spectrum. Depending on the value of $\alpha$, such signals correspond
to well-defined physical processes [white noise ($\alpha =0$), $1/f$
  noise ($\alpha =1$), random walk ($\alpha =2$), random acceleration
  ($\alpha =4$), etc.].  As $\alpha$ increases, the signal changes
from uncorrelated ($\alpha =0$) to weakly correlated ($0<\alpha <1$),
and then strongly correlated for $1\le \alpha \le\infty$.  Thus, the
characterization of correlations is straightforward, and the
distribution of the maximum, $x_1={\rm max}_tx(t) -\overline {x(t)}$,
measured with respect to the time average, $\overline {x(t)}$, has
been discussed in the
literature\cite{Majum2004,Gyorgyi2007,Fjodorov2009}.  The present work
is an extention of Ref.\cite{Gyorgyi2007} to order statistics.

Before summarizing the results, a clarification is in order.  Namely,
the signal $x(t)$ is continuous for $\alpha >1$ and thus, while the
meaning of the maximum $x_1$ is clear, the definition of $x_2$ (the
$2^{nd}$ largest) etc. may not be obvious. The key here is to
recognize that, as discussed in Sec.II., the signal is determined
through a finite number of $N$ Fourier amplitudes. The $N$
independent values of the signal can be obtained e.g. by sampling at
equal intervals of $T/N$, yielding a discretized signal in which
the $2^{nd}$, $3^{rd}$ largest etc. are well-defined.
 
Our main result concerns the average gap 
$d_k=\langle x_k-x_{k+1}\rangle$ between the $k^{th}$ and $(k+1)^{st}$ 
largest. We find that it scales with $k$ as
\be
d_k\sim 
\begin{cases} 
\, k^{-1} & \text{ for $0\le \alpha <1$}\, ,\\
\, k^{(\alpha -3)/2} &\text{ for $1<\alpha < 5$}\, ,\\
\, k & \text{ for $5 < \alpha \le \infty$}\, .
\end{cases}
\label{gapscaling}
\ee
The above results are first obtained from 
simulations of the $1/f^\alpha$ signals and then phenomenological 
arguments are also used to derive the exponents. 
As can be seen, there are three regimes and 
the scaling exponents match at the borderline values $\alpha =1$ and $5$.
It is suspected, however, 
that the power laws at $\alpha =1$ and $5$ have logarithmic corrections 
which are not resolved by the present simulations.

The gap scaling \eqref{gapscaling} implies that the spectrum of the
average values of the $k^{th}$ largest behaves as \be
\varepsilon_k=\langle x_1-x_k \rangle \sim
\begin{cases} 
\, \ln k & \text{ for $0\le \alpha <1$}\, ,\\
\, k^{(\alpha -1)/2} &\text{ for $1<\alpha < 5$}\, ,\\
\, k^2 & \text{ for $5 < \alpha \le \infty$}\, .
\end{cases}
\label{spectrum}
\ee

The above spectrum for $1<\alpha\le\infty$ can also be obtained 
by relating $\varepsilon_k$ to the density
of near extreme states which, for periodic signals, is known to be 
given by the distribution of the maxima of the signal relative 
to the initial value \cite{Burkhardt}. 
This distribution has been studied in detail
in \cite{Burkhardt} and, in particular, the $\alpha =2$, $4$, and
$\infty$ cases were solved exactly. Thus, provided the assumption 
about the scaling form 
$\varepsilon \sim k^\beta$ is valid, the exponents for 
$\alpha =2$, $4$, and $\infty$ cases in \eqref{spectrum} are exact. 

Interestingly, the same spectrum \eqref{spectrum} emerges
in the quasi-classical limit of a quantum mechanical system. Namely, 
the energy spectrum $\varepsilon_k$ of a particle in a one-dimensional 
potential $U(z)\sim |z|^\theta$ is given by $\varepsilon_k\sim
k^{2\theta/(2+\theta)}$.  Thus, there is a one-to-one correspondence
between the exponents of the order statistics spectrum for $1<
\alpha \le \infty$ and of the quantum mechanical energy spectrum for
$0< \theta \le \infty$.

In order to arrive at the above results, we begin with a short 
introduction of $1/f^\alpha$ signals (Sec.~\ref{1falpha})
followed by the description of the simplest case of $\alpha =0$
(i.i.d. variables) in Sec.~\ref{alphanull}. 
Numerical simulations together with scaling considerations 
for general $\alpha$ are discussed in Sec.~\ref{numerics}. 
The results are rederived from the
point of view of density of near extreme states 
in Sec.~\ref{nearextreme}, and the relationship to 
quasi-classical quantum spectra is presented in Sec.~\ref{Qspectra}.

\section{Periodic, Gaussian $1/f^\alpha$ signals}
\label{1falpha}

Detailed discussions of periodic, Gaussian $1/f^\alpha$ signals and
their correlation and roughness properties can be found in
Refs.\cite{Gyorgyi2007,Antal2002}.  Here, we just define the relevant
notation and describe how the order statistics is evaluated
numerically.

The configurational weight of a Gaussian $1/f^\alpha$ signal $x(t)$ of 
periodicity $T$ is given by
\begin{equation}
  \mathcal{P} [x(t)] \propto e^{-S[x(t)]}\;,\label{weight}
\end{equation}
where the effective action 
\begin{equation}
  S[\{c_n\};\alpha] = (2\pi)^{\alpha} T^{1-\alpha} \sum_{n=1}^{N/2}
  n^{\alpha}\label{Fourieraction}
  |c_n|^2\; ,
\end{equation}
is defined through the Fourier coefficients, $c_n$, of the signal 
\begin{equation}
x(t) = \sum_{n=-N/2+1}^{N/2} c_n e^{2\pi i nt/T}\;,\quad
c_n^*=c_{-n}\;.\label{Fourierseries}
\end{equation}
Here $N$ is a positive, even integer, and the phase of $c_n$ is drawn
randomly and uniformly from the interval $[0,2\pi]$.

From Eqs. (\ref{weight}) and (\ref{Fourieraction}) one sees that the
amplitudes of the Fourier modes are independent, Gaussian distributed
variables, but only for $\alpha=0$ are they identically
distributed. This is also apparent from the mean square amplitude
$\langle |c_n|^2 \rangle \propto 1/n^{\alpha}$ which is consistent
with a $1/f^\alpha$ power spectrum and independent of $n$ only for
$\alpha=0$.  Although the Fourier components $c_n$ are uncorrelated, the
corresponding time signal $x(t)$ is correlated at different times $t$
and $t'$ for $\alpha>0$, and the correlation increases with increasing
$\alpha$. In particular, the correlation function 
$\langle x(t')x(t'+t)\rangle$ is bounded for $0 \le
\alpha < 1$ (weakly correlated regime), while it diverges 
for $\alpha \ge 1$ in the 
limit $T\to\infty$ with $t/T$ finite (regime of strong correlations) \cite{Gyorgyi2007}.

The action in Eqs. (\ref{weight}) and
(\ref{Fourieraction}) may be formally written in the continuum limit as
\begin{equation}
S[x(t)]=\frac{1}{2}\int_0^T dt\thinspace\left\vert\frac{d^{\alpha/2}x}
{dt^{\alpha/2}}\right\vert^2 \;,\label{realspaceaction}
\end{equation}
implying the stochastic equation of motion
\begin{equation}
\frac{{d^{\alpha/2}x}}{dt^{\alpha/2}}=\xi(t)\;,\quad
\langle\xi(t)\xi(t')\rangle=
\delta(t-t')\;,\label{stochasticeq}
\end{equation}
where $\xi(t)$ is Gaussian white noise with zero mean. In this form it
is transparent that the signal for $\alpha=0$, $2$, and $4$, describes
white noise, random walk, and random acceleration, respectively.

It is important to note that since the maximum frequency
appearing in the sum \eqref{Fourierseries} is $N/T$, the series does 
not resolve fine structure on time scales less than $\tau = T/N$. 
Thus, we may view the signal $x(t)$ as a batch of $N$ variables
$x_n=x(nT/N)$ ($n=0,1,...,N-1$), and determine the order statistics
by ordering the $x_n$. As usual in extreme statistics,
we should consider the limit of the batch size going to infinity
($N\to\infty$) and search for finite results after appropriate 
rescalings. In our case, the limit of the batch size going to 
infinity ($N\to\infty$) is equivalent to taking the  
limit $T\to \infty$ with $\tau=T/N$ kept fixed.

\begin{figure*}[htb]
\begin{center}
\includegraphics*[width=\columnwidth]{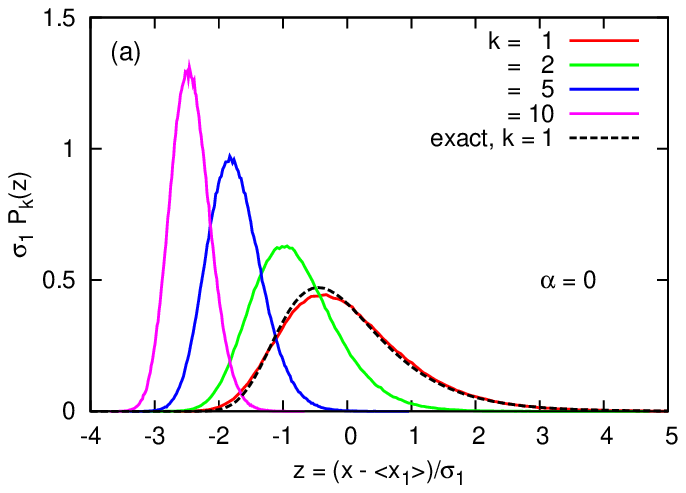}
\includegraphics*[width=\columnwidth]{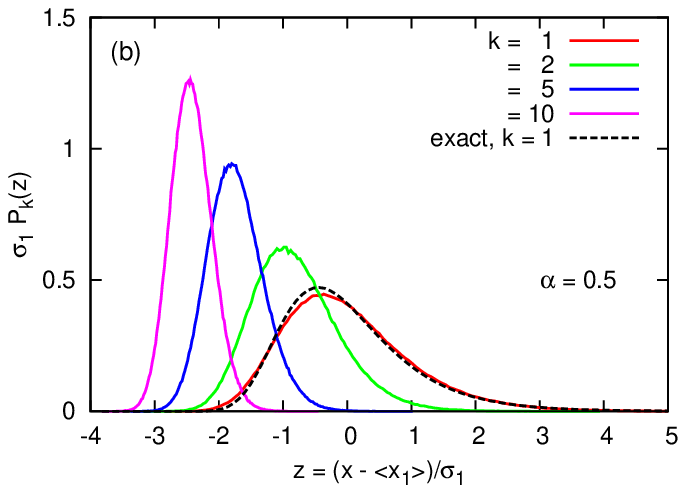}\\
\includegraphics*[width=\columnwidth]{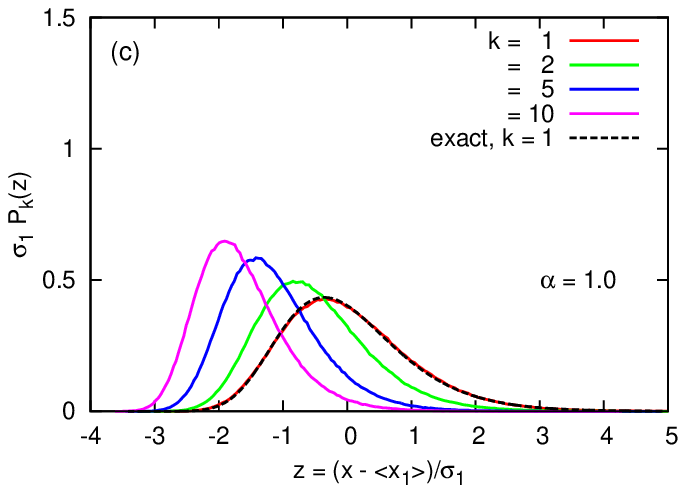}
\includegraphics*[width=\columnwidth]{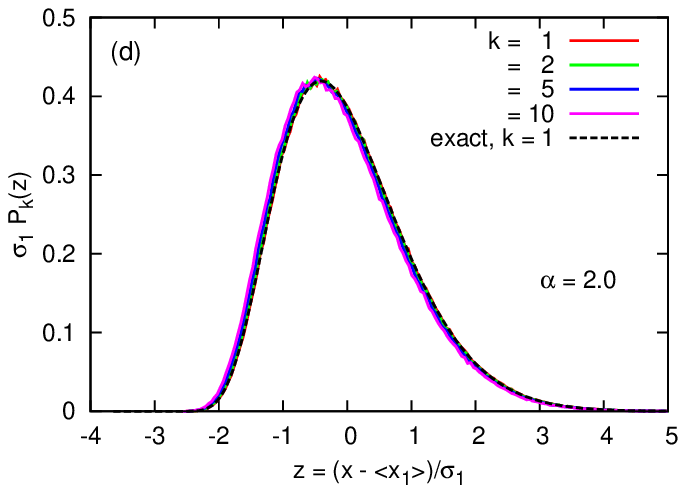}
\end{center}
\caption{(Color online) Probability distribution of the $k = 1,2,5,10$
  maxima (right to left) for (a) $\alpha = 0$, (b) $\alpha = 0.5$, (c)
  $\alpha = 1$, and (d) $\alpha = 2$, centered and scaled according to
  the mean and standard deviation of the $k=1$ maximum. $N = 16\,384$
  in all cases and the number of simulations ($10^6$) is large enough
  so that the error bars are of the order of the width of the lines
  drawn. The exact limit distributions are also plotted for $x_1$. One
  can see that the finite size corrections are large for $\alpha <1$.
\label{F:histograms}}
\end{figure*}

We now describe how the order statistics is evaluated numerically.
For a given $\alpha$ and $N$, the amplitude and phase of the Fourier
coefficients in \eqref{Fourierseries} are sampled from their (Gaussian
and homogeneous, respectively) distributions, and the signal $x(t)$ is
generated by fast Fourier transform. The values $x_n=x(nT/N)$ are
ordered and the largest $x_1$, $2^{nd}$ largest $x_2$, etc.  are
determined. The process is repeated $10^6$ times in order to obtain
the distribution $P_N(x_k)$, and to have well defined averages
$\langle x_k\rangle_N$, as well as gaps $d_{k,N}=\langle
x_k\rangle_N-\langle x_{k+1}\rangle_N$. Signals are generated in such
a way for $N=16,32,...,16384$, and the $N$-dependence of the above
quantities is examined. Finally, appropriately scaled quantities are
introduced, so that a finite structure emerges in the $N\to\infty$
limit.

\section{Order statistics for $\alpha =0$}
\label{alphanull}

We begin with the $\alpha =0$ limit, with $N$ i.i.d. variables drawn
from a Gaussian parent distribution.  The results for order statistics
are known \cite{Leadbetter} in this case.  In particular, the $k^{th}$
maximum $\langle x_k \rangle$ and its root-mean-square fluctuation
$\sigma_k=\sqrt{\langle x_k^2 \rangle-\langle x_k\rangle^2}$ scale for
large $N$ as
\be \langle
x_k \rangle \sim \sqrt{\ln{N}}\quad , \quad \sigma_k\sim
1/\sqrt{\ln{N}} \, .
\label{Gaussscale}
\ee
The above scalings suggest that, in the $N\to\infty$ limit, 
finite distribution functions are obtained 
by introducing the scaled variable $z$ through $x=a_Nz+b_N$, where 
$a_N\sim 1/\sqrt{\ln{N}}$ and $b_N\sim \sqrt{\ln{N}}$. Indeed,
using the ``experimental" scaling $z=(x-\langle x_1\rangle )/\sigma_1$,
the limit distribution $P_k(z)$ for the $k^{th}$ maximum 
is given by \cite{Leadbetter}
\be
P_k(z)=\frac{1}{(k-1)!}\exp{(-k\tilde z-e^{-\tilde z})}\,\,\,\, , \,\,\,\,  
\tilde z =az +\gamma 
\label{general-kth}
\ee
where $a=\pi/\sqrt{6}$ and $\gamma$ is Euler's constant. 

The $P_k(z)$ distribution functions obtained from simulations of
$N=16\,384$ independent modes are displayed in
Fig.\ref{F:histograms}(a) for $k=1$ [Fisher-Tippett-Gumbel (FTG)
  distribution], $k=2$, $5$, and $10$. As one can see, even
$N=16\,384$ is not large enough to arrive at the limit
distributions for $\alpha <1$. 
This is related to the notoriously slow (logarithmic)
convergence, as discussed in detail in \cite{Gyorgyi2007,Gyorgyi2008}.
 
Using \eqref{general-kth} to express $\langle z_k \rangle$ via
$\langle z_{k+1} \rangle$, the gap between the $k^{th}$ and
$(k+1)^{st}$ largest is given by
\be
d_k=\langle z_k \rangle-\langle z_{k+1} \rangle = \frac{1}{ak} \,.
\label{iidkth}
\ee
The spectrum of the average values $z_k$ can be evaluated as the 
sum of the gaps, and one finds for large $k$ 
\be
\varepsilon_k\equiv\langle z_1 \rangle-\langle z_k \rangle = 
\sum_{\ell =1}^{k-1}d_\ell=
\frac{1}{a}\sum_{\ell =1}^{k-1} \frac{1}{\ell}\approx\frac{1}{a}\ln{k} \,.
\label{iidepsk}
\ee
One should note here that both the distribution $P_k(z)$ and the
results for the gap and the spectrum remain valid for all parent
distributions whose tail extends to infinity and decay faster than any
power law (domain of attraction for the FTG universality class).
 
In the rest of the paper, we shall consider the case $\alpha \not= 0$
and concentrate on $P_k(z)$, $d_k$, and $\varepsilon_k$. In explaining
the simulation results, we shall make use of a phenomenological
argument for calculating $d_k$, which we demonstrate now for the
i.i.d. case. 

Let us assume that we are studying the order statistics in the
unrescaled variable ($x_1,x_2,...$), and we discover from simulations
that $\langle x_1 \rangle\sim \sqrt{\ln{N}}$ and $\langle
x_k-x_{k+1}\rangle \sim d_k/\sqrt{\ln{N}}$. Summing over $N/2$ gaps
gives the distance from the largest to the 
to the median of the parent distribution 
[$\langle x_{N/2}\rangle \sim O(1)$], such that
\be
\sum_{k=1}^{N/2}\langle x_k-x_{k+1}\rangle \approx 
\sum_{k=1}^{N/2} d_k/\sqrt{\ln{N}} \approx
\langle x_1\rangle \approx \sqrt{\ln{N}}\,.
\label{pheno1}
\ee
Multiplying by $\sqrt{\ln{N}}$, the above equalities yield
\be
\sum_{k=1}^{N/2} d_k\approx \ln{N}\,.
\label{pheno2}
\ee
Assuming that the above sum is dominated by the $d_k$ with simple 
power-law behavior $d_k\sim k^{-\delta}$, one can read off the 
scaling of the gap
\be
 d_k\sim k^{-1}\,
\label{pheno3}
\ee
in agreement with the exact expression \eqref{iidkth}.

The scaling assumption, backed up by simulations, can thus be used to
develop relationships between observed quantities.

\section{Order statistics for $\alpha \not= 0$}
\label{numerics}

For $\alpha\not= 0$, we shall present the numerical results in the
following order: First, the probability distribution $P_k$
(Fig.\ref{F:histograms}),  second, the $N$-dependent scaling of
$\langle x_1\rangle$, $\sigma_1$, and $\langle x_1-x_2\rangle$
(Fig.\ref{F:scaled_gaps}) are discussed.  
Then, the
average gap $d_k=\langle x_k-x_{k+1}\rangle$ as a function of $k$
(Fig.\ref{F:dk}) is described with the $N$-scaling removed  
when necessary. Finally,
the spectrum $\varepsilon_k$ is obtained by summing up the $d_k$.

The numerical results indicate that, depending on $\alpha$, 
there are three distinct regimes of order statistics as we 
shall discuss below. 

\subsection{Weakly correlated regime ($0< \alpha <1$)}
\label{alpha<1}

The distributions $P_k(z)$ obtained for $\alpha =0.5$
[Fig.\ref{F:histograms}(b)] are characteristic of the results. As can
be seen, the distributions $P_k(z)$ in (b) are rather close to those
in (a). In fact, the limit distribution of the maximum $x_1$, in the
interval $0\le \alpha <1$, is known to be FTG distributed just as in
the i.i.d. case \cite{Berman1964}. At the same time, however, the
finite-size corrections are large in this regime \cite{Gyorgyi2007},
and it is not surprising that the deviations from the limit
distribution are significant even for $N=16\,384$.

\begin{figure}[htb]
\includegraphics*[width=\columnwidth]{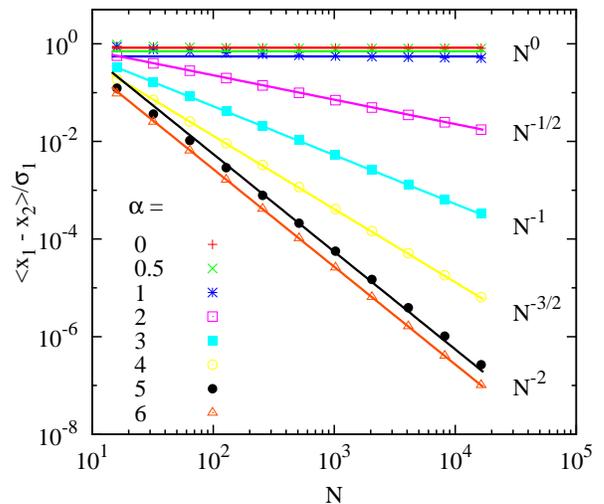}
\caption{(Color online) 
  Difference between the $k = 1 $ and $k = 2$
  average maxima as a function of $N$ for increasing $\alpha$ (top to
  bottom). The asymptotes are indicated on the right.\label{F:scaled_gaps}}
\end{figure}

Figs.\ref{F:histograms}(a),(b) reveal that the widths $\sigma_k$ of
the distributions $P_k(x)$ are of the same order as those for $P_1(x)$
and, since it is known that $\sigma_1\sim 1/\sqrt{\ln N}$, we have
$\sigma_k\sim 1/\sqrt{\ln N}$.  Furthermore, the gaps $\langle
x_k-x_{k+1}\rangle$ are also of the same order as $\sigma_1$ and so
$\langle x_k-x_{k+1}\rangle\sim 1/\sqrt{\ln N}$.  This is more
precisely checked in Fig.\ref{F:scaled_gaps}, where the ratio $\langle
x_1-x_2\rangle/\sigma_1$ is plotted and, indeed, is of the order
${\cal O}(1)$. Similar results are obtained for all the $\langle
x_k-x_{k+1}\rangle/\sigma_1$ ratios we examined ($k\le 10$).

\begin{figure}[htb]
\includegraphics*[width=\columnwidth]{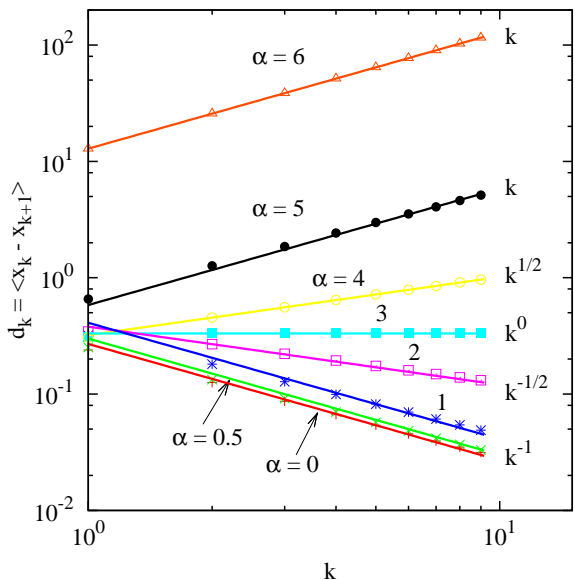}
\caption{(Color online) 
  Average unscaled gap $d_k$ for $\alpha = 0,0.5,1,\ldots,6$ 
  (bottom to top) obtained from simulations with $N=16384$. The
  values of the limiting slopes are indicated at the end of the lines. 
  The straight line fits are excellent except for $\alpha =1$ and $5$ 
  where logaritmic corrections may be present. 
\label{F:dk}}
\end{figure}

Once it is established that $\langle x_k-x_{k+1}\rangle\sim
1/\sqrt{\ln N}$, we can plot $d_k=\langle
x_k-x_{k+1}\rangle\sqrt{\ln{N}}$ as a function of $k$ to examine the
scaling properties of the gap $d_k$.  As Fig.\ref{F:dk} shows,
$d_k\sim k^{-1}$, in agreement with the expectation that the weak
correlations in the $0<\alpha <1$ regime do not affect the extreme
statistics properties of the signal as compared to the i.i.d. case.

It should be noted that the $\langle x_k-x_{k+1}\rangle\approx
d_k/\sqrt{\ln{N}}$ scaling in conjunction with the mathematical result
$\langle x_1\rangle \sim \sqrt{\ln N}$ allows one to repeat the
phenomenological argument given in Sec.\ref{alphanull} [see
  eqs.\eqref{pheno1}-\eqref{pheno3}] to establish that $d_k\sim
k^{-1}$ and $\varepsilon_k\sim \ln{k}$ \eqref{iidepsk}, valid in the
large $k$ limit.

\subsection{Regime of non-trivial scaling ($1\le \alpha <5$)}
\label{alpha<5}

For $\alpha \ge 1$, we enter the strongly correlated regime where
fluctuations diverge with system size. The large-fluctuation regime
can be further divided according to whether the second derivative of
the signal (playing an important role in determining the order
statistics) is continuous ($5 \le \alpha \le \infty$) or not ($1 \le
\alpha < 5$).  Sec.\ref{alpha<5} will be devoted to $1 \le \alpha <
5$.

We begin with the borderline case ($\alpha = 1$) separating the weakly
and strongly correlated regimes. The histograms are plotted in
Fig.\ref{F:histograms}(c) and, compared to $\alpha <1$, they draw
closer to each other both in terms of location and scale.  Since the
$\alpha = 1$ case lies at the threshold between the two different
scaling regimes for $\alpha<1$ and $\alpha >1$, the full quantitative
details are difficult to extract. There are, however, some exact
results for $x_1$: the first maxima scales with $N$ as $\langle x_1
\rangle \sim \ln N$ and, furthermore, the probability distribution
$P(y)$ for the shifted maximum $y=x_1-\langle x_1 \rangle$ is given by
\cite{Fjodorov2009}
\be
P(y)=[2e^{y/2}K_1(2e^{y/2})]^\prime
\ee
where $K_1$ is the modified Bessel function. The agreement between
$P(y)$ and numerics in Fig.\ref{F:histograms}(c) is excellent.

As for the width of the distributions and the scaling of the gaps for
$\alpha=1$, we only have indications from simulations. It appears that
$\sigma_{k>1} \sim O(1)$, as for $k=1$, and that the gaps $d_k$ are
also $O(1)$ for all examined $k$. Logarithmic crossovers prevent us
from making any firm conclusions.

For $\alpha >1$, the convergence to the limit distribution becomes
power law as seen in extensive simulations for general $\alpha$
\cite{Gyorgyi2007} and also in the exact solution for $\alpha=2$
\cite{Majum2004,MajumFSSSchehr}.  Thus the convergence is faster and
the order statistics can be reliably investigated for $\alpha>1$ (but
not too close to $\alpha =1$).

\begin{figure}[htb]
\includegraphics*[width=\columnwidth]{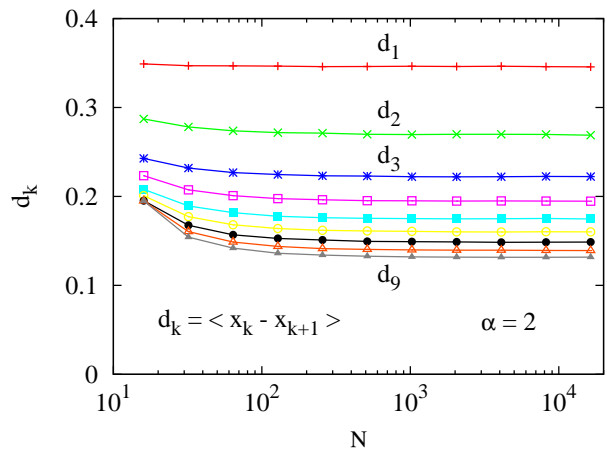}
\caption{(Color online) Unscaled gaps for $\alpha =2$. 
The simulation results are for $N=16,32,...,16\,384$, and the lines are just 
guides for the eyes.
\label{F:a2-dk}}
\end{figure}

Fig.\ref{F:histograms}(d) displays the $\alpha =2$ results for $P_k(z)$,
in which one can observe a new scaling regime. Namely, the histograms
are barely distinguishable, i.e. the widths of the histograms
$\sigma_k$ are much larger than the gaps $d_k$. It is known
\cite{Majum2004} from the exact solution for $P_1(x)$ for $\alpha =2$
that $\sigma_1\sim \langle x_1 \rangle \sim N^{1/2}$. Thus our
simulation results (see Fig.\ref{F:scaled_gaps}) indicating $\langle
x_1-x_2\rangle/\sigma_1\sim N^{-1/2}$ confirm that the first gap
$d_1=\langle x_1-x_2\rangle $ is indeed $\mathcal{O}(1)$. It then
follows that there is no need for an $N$-dependent rescaling of the
gap.  The same result is obtained for all $d_k=\langle
x_k-x_{k+1}\rangle$ as seen directly in Fig.\ref{F:a2-dk}, where the
unscaled gaps for $\alpha =2$ are plotted as functions of $N$.

Plots similar to those in Fig.\ref{F:a2-dk} implying that $d_k\sim
O(1)$ in the large $N$ limit can be produced for any $1<\alpha <5$.
The same conclusion can also be drawn from Fig.\ref{F:scaled_gaps}.
Indeed, it is known \cite{Gyorgyi2007} that $\sigma_1\sim
N^{(\alpha-1)/2}$, and since the asymptotes of the simulation results
for $d_k/\sigma_1$ scale like $N^{-(\alpha-1)/2}$, one concludes that
$d_k\sim O(1)$.

Once $d_k\sim O(1)$ is established, we can plot the unscaled $d_k$ as
a function of $k$. As can be seen in Fig.\ref{F:dk}, the $d_k$ display
power law scaling and the exponents for $\alpha = 2,\,3,\,4$ are
$-1/2,0$, and $1/2$, respectively.  These exponents, together with the
$-1$ and $+1$ values (with possible logarithmic corrections) at
$\alpha =1$ and $5$, suggest that in the interval $1< \alpha <5$
\be
d_k\sim k^{(\alpha-3)/2} \, .
\label{d_k-scaling}
\ee 
The above scaling can be derived from the phenomenological considerations
embodied in eqs.\eqref{pheno1}-\eqref{pheno3} which worked for $\alpha <1$.
The starting point is the observation that the sum of $d_k$ is of the 
order of $x_1$, and since $x_1 \sim N^{(\alpha-1)/2}$, we can write 
\be
\sum_{k=1}^{N/2} d_k\approx \langle x_1 \rangle \sim N^{(\alpha-1)/2} 
\ee
Assuming that the $d_k\sim k^\delta$ scaling is valid for large
$k$ of the order of $N$, we have
\be
\sum_{k=1}^{N/2} k^\delta \approx N^{(\alpha-1)/2} \,
\ee
and the above equality implies $d_k\sim k^{(\alpha-3)/2}$, as suggested by 
simulations.

\begin{figure}[htb]
\includegraphics*[width=\columnwidth]{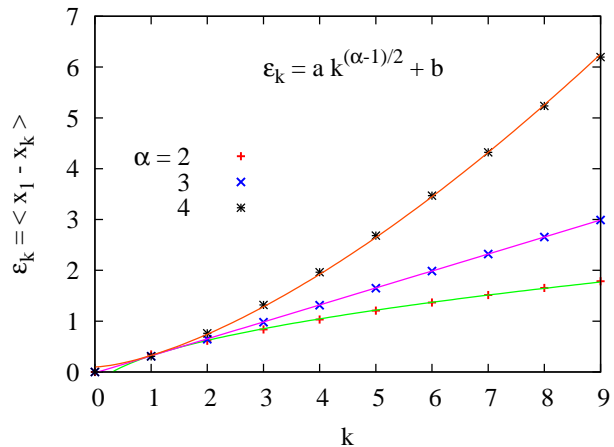}
\caption{(Color online) Spectra of the ordered values $\langle x_k
  \rangle$ measured from the largest value of the signal
  $\varepsilon_k =\langle x_1 -x_k \rangle$ for $\alpha =2$, $3$, and
  $4$. Simulation results for $N=16\,384$ are plotted together with
  two-parameter fits (solid lines) of the form $\varepsilon_k
  =ak^{(\alpha -1)/2} +b$ where $a$ and $b$ depend on $\alpha$.
\label{F:epsk-a234}}
\end{figure}

The large $k$ asymptote of the spectrum $\varepsilon_k$ can now be 
obtained by integrating the $d_k\sim k^{(\alpha-3)/2}$ expression, yielding
\be
\varepsilon_k\sim  k^{(\alpha-1)/2} \, .
\label{eps_kalpha}
\ee
As can be seen from Fig.\ref{F:epsk-a234}, fits of the form
$\varepsilon_k=ak^{(\alpha-1)/2} +b$ are indeed excellent for $\alpha
=2,3,4$. It should be noted, however, that the constant $b$ is
important for observing the exponent $(\alpha-1)/2$. Naive
straight-line fits on log-log plots of $\varepsilon_k$ versus $k$
yield poor estimates for the exponents since the constant $b$
generates strong corrections to scaling in the range $0<k<10$ studied.


\subsection{Large $\alpha$ regime ($5\le \alpha \le\infty$)}
\label{largea}

The large $\alpha$ regime is special in that the second derivative of
the signal becomes continuous for $\alpha \ge 5$.  Then the largest
local maximum $x_1=x(t_1)$ and the values near it are expected to
determine the order statistics, and this allows one to calculate the
spectrum by expanding the signal around the maximum
\be
x(t_1+\Delta)=x_1+\frac{1}{2}x^{\prime\prime}(t_1)\Delta^2+O(\Delta^3)\,.
\label{expansion}
\ee 
where $\Delta =\tau$, $2\tau$, ..., $(k-1)\tau$ are the positions of
the second, third, ..., $k$th maximum respectively. The justification
for the above expansion comes from comparing the first and second
terms on the right hand side of \eqref{expansion}. The maximum value
$x_1$ scales as $\langle x_1\rangle \sim N^{(\alpha-1)/2}$, and the
characteristic value of $x^{\prime\prime}$ can be estimated through
$[{\langle {(x^{\prime\prime}})^2 \rangle}]^{1/2}$. A straightforward
calculation yiels
\be
[{\langle {(x^{\prime\prime}})^2 \rangle}]^{1/2}
\sim \langle |x^{\prime\prime}(t_1)|\rangle \sim N^{(\alpha-5)/2} 
\sim \langle x_1\rangle N^{-2} \, 
\label{xprimeprime}
\ee
showing that the second term is indeed much smaller than the first one
(note that $\Delta\sim \tau=T/N\sim O(1)$ in the limit we are considering). 

Using the expansion (\ref{expansion}), 
the unscaled gap $d_k$ can be written as
\be
\langle x(t_1+(k-1)\tau)-x(t_1+k\tau)\rangle
\approx \langle |x^{\prime\prime}(t_1)|\rangle \tau^2k \, .
\label{large-alpha}
\ee
Next, the $N$-dependence contained in 
$\langle |x^{\prime\prime}(t_1)|\rangle\tau^2\sim N^{(\alpha-5)/2}$ 
is scaled out, and we obtain the gap as a function of $k$
\be
d_k \sim k\, 
\label{dk-a5}
\ee
and the corresponding spectrum then follows as
\be
\varepsilon_k \sim k^2\,.
\label{ek-d5}
\ee
As can be seen in Fig.\ref{F:dk}, the $\alpha =5$ and $6$ results for $d_k$
are in excellent agreement with the $d_k \sim k$ scaling for small values of
$k$ as well.

It should be noted that the scaled first gap shown for $\alpha = 5$
and $6$ on Fig.\ref{F:scaled_gaps} can be also understood on the basis
of the expansion (\ref{expansion}-\ref{large-alpha}). Indeed, noting
that $\langle x_1\rangle \sim \sigma_1$, and using \eqref{xprimeprime}
with $k=1$, we have
\be
\frac{\langle x(t_1)-x(t_1+\tau)\rangle}{\sigma_1}
\sim 
\frac{\langle |x^{\prime\prime}(t_1)|\rangle }{\sigma_1} \sim \frac{1}{N^2}\, ,
\label{scgap-a5}
\ee
in agreement with the $\alpha \ge 5$ results on Fig.\ref{F:scaled_gaps}.
The above result is exact in the $\alpha\to\infty$ limit where 
the signal is just a single mode $x(t)\sim x_1\sin{(2\pi t/T)}$, 
thus the relationships $\langle x_1\rangle \sim \sigma_1$ and 
$\langle|x^{\prime\prime}(t_1)|\rangle \sim \langle x_1\rangle/N^2$
are trivially satisfied.

The simulations for $\alpha=5$ and $6$, together with the simplicity
of the $\alpha\to\infty$ limit, suggest that the results
(\ref{dk-a5}-\ref{scgap-a5}) are valid for all $\alpha\ge 5$. The
possible logarithmic corrections at the borderline point $\alpha=5$
are outside the limits of our present simulations.

\section{Summary and a connection to the density of 
near extreme states}
\label{nearextreme}

The results for the gap $d_k$ and the spectra $\varepsilon_k$ obtained
in Sec.\ref{numerics} are summarized in columns 1-3 of Table
\ref{Table:spectrum}.  As one can see from columns 2 and 3, there are
three regimes and, furthermore, the gaps and the spectrum change with
the noise properties ($\alpha$) only in the $1\le \alpha < 5$ regime.

\begin{table}[htb!]
\caption{\label{Table:spectrum} Gaps, $d_k$ and spectra $\varepsilon_k$ 
in the order statistics of $1/f^\alpha$ signals as obtained in simulations
and suggested by scaling arguments. The fourth column displays the 
one-dimensional potential $V(x)$ in which the quantum mechanical motion of
a particle generates energy spectra with the same scaling properties.}   
\begin{ruledtabular}
\begin{tabular}{lccc}
\; \; \; $\alpha$ & $d_k$  & $\varepsilon_k$  & $V(x)$ \\[5pt]
\hline\noalign{\medskip}
$0\le\alpha <1$ & $k^{-1}$ & $\ln k$ & $\ln|x|$ \\[5pt]
\hline\noalign{\medskip}
$1\le\alpha <5$  & $k^{(\alpha-3)/2}$ & $k^{(\alpha -1)/2}$  & $|x|^{2(\alpha -1)/(5-\alpha)}$\\[5pt]
\hline\noalign{\medskip}
$5\le \alpha \le\infty$  & $k$ & $k^2$ &   $|x|^\infty$ \\[3pt]
\end{tabular} 
\end{ruledtabular}  
\label{Table1}
\end{table} 

The above results for $1<\alpha\le\infty$ can be derived by
considerations related to the density of near extreme states. Namely,
if we assume that the spectrum has a scaling form $\langle
x_1-x_k\rangle=\varepsilon_k \sim k^\beta$ then the density of states
near the extreme is $\rho (\delta x) \sim (\delta x)^{1/\beta}/\delta
x\sim (\delta x)^{1/\beta -1}$, thus knowledge of the small argument
asymptote $\rho (\delta x) \sim (\delta x)^\gamma$ allows one to
deduce the exponent $\beta =1/(\gamma +1)$.

The density of near extreme states $\rho(\delta x)$ was first
investigated for i.i.d. 
variables by Sabhapandit and Majumdar \cite{Sabha-Majum}. Later,
it was shown by Burkhardt et al. \cite{Burkhardt} 
that $\rho(\delta x)$ can be obtained 
for periodic signals as the distribution $\Phi_I(x)$ of the maximum  
with respect to the initial value $x={\rm max}_tx(t)-x(0)$. The 
distribution $\Phi_I(x)$ for periodic $1/f^\alpha$ signals has been 
investigated by simulations as well as through exact solutions
for the particular cases of $\alpha =2$, $4$, and $\infty$ 
\cite{Burkhardt}.
The small argument behavior of $\Phi_I$ was indeed found to have 
a power law form
$\Phi_I(\delta x)\sim (\delta x)^\gamma$ with 
[\,see eq.(65) in \cite{Burkhardt}\,]
\begin{equation}
\gamma(\alpha)=\begin{cases} \displaystyle\frac{3-\alpha}{\alpha -1}\;,
 &\alpha < 5\;,\\[10pt]\displaystyle
-\frac{1}{2}\;, &\alpha \ge 5 \;. \end{cases}
\label{gammaalpha}
\end{equation}
The scaling exponents of the spectra following from the above
expression [\,$\beta =1/(\gamma +1)$\,] are equal to those in Table
\ref{Table:spectrum} for $1< \alpha \le\infty$, thus reinforcing our
earlier conclusions. Since the exponents in \eqref{gammaalpha} are
exact for $\alpha =2$, $4$, and $\infty$, we can conclude that the
corresponding $\beta(\alpha)$ exponents depend on the validity of the
$\varepsilon_k\sim k^\beta$ scaling.

\section{Comparing with quantum spectra}
\label{Qspectra}

We shall now compare the order statistics spectra to the energy
spectra of quantum mechanical systems in the quasi-classical
limit. The reason for this comparison, apart from its entertaining
aspects, is that the discrete quantum mechanical spectra may also be
considered as an order statistics spectra.

Let us consider a particle of mass $m$ which moves in a potential
\be
V(x)=g|x|^\theta
\ee
where $g>0$ is the coupling constant. The simplest way to calculate
the quasi-classical limit of the spectra is to use dimensional
analysis combined with the observation that, in the large
quantum-number limit ($k\to\infty$), the quantization condition ($\int
\! pdq=kh$) forces $h$ and $k$ to appear in the combination $hk$.
This means that the $k$-dependence of the spectra is determined by its
$h$ dependence.  Since the energy is uniquely determined by the
dimensions of $m$, $q$, and $h$, one obtains
\be
E_k\sim (hk)^\frac{2\theta}{\theta +2} \, .
\label{qmspectrum}
\ee

Comparing the above result with Column 3 in Table \ref{Table1}, 
one can see that there is a mapping between the quantum mechanical and
the order statistics exponents. The large-$\alpha$ regime 
($5\le \alpha \le \infty$) corresponds to the infinite square-well
potential ($\theta =\infty$), while $\theta$ changes 
monotonically from $0$ to $\infty$ in the nontrivial regime ($1< 
\alpha <5$), and the correspondence is given by
\be
\theta =\frac{2(\alpha -1)}{5-\alpha}\, .
\ee
The above considerations cannot be applied for the 
remaining $0\le \alpha \le 1$ range but the  
order statistics spectrum $\varepsilon_k\sim \ln k$ suggests 
that the corresponding potential is of the form $V(x)\sim \ln |x|$. 

We would like to emphasize that it is not obvious that the mapping
between the exponents have any physical content. Nevertheless, it is
intriguing to ask whether there is a {\it quasi-classical} extreme
value question whose answer is the quantum mechanical spectra.

\section{Final remarks}

Studying the order statistics in $1/f^{\alpha}$ signals, we found 
three scaling regimes as summarized in Table \ref{Table1}. 
In the case of weakly correlated stationary signals ($0 <
\alpha < 1$), the i.i.d. result ($d_k\sim k^{-1}$) applies, 
while for strongly correlated
signals ($1 \le \alpha \le \infty$) the scaling depends on whether the
signal is twice differentiable ($\alpha \ge 5$) or not ($1 \le \alpha
< 5$). In the former case, order statistics ($d_k\sim k$) follows 
from expanding the signal around the maximum. 
In the latter case, the observed
scaling ($d_k\sim k^{(\alpha -3)/2}$) 
is derived using a phenomenological argument (exact for
i.i.d. variables), namely that the
sum over $d_k$ scales in the same way as $\langle x_1 \rangle$. 
Meanwhile, the spectrum $\varepsilon_k$ is essentially obtained
by integrating $d_k$. The same scaling picture was also obtained 
by relating order statistics to the density of near extreme 
states studied in previous works.

It is clear that investigating the order statistics of correlated
systems will help in characterizing and understanding  
extreme events in more detail. It is also clear, however, that  
the results for $1/f^\alpha$ signals have a restricted range 
of applicability, and much more is needed
to advance our understanding of the effects of various correlations.
Nevertheless, there are phenomena where $1/f^\alpha$ type 
fluctuations do emerge, the extremes are important and, consequently, 
our results may be utilized. A possible application is to
climate records (temperature, precipitation, etc.) which 
often display e.g. clustering of extreme events, and the 
power spectrum of the fluctuations appears to be of $1/f^\alpha$ 
type \cite{Bunde2006,Bunde1}.

\begin{acknowledgments}
This research has been supported by the 
Hungarian Academy of Sciences through OTKA Grants
Nos.\ K 68109 and NK 72037. We would like to thank
T. Burkhardt, S. N. Evans, G. Gy\"orgyi, and M. Z. R\'acz for 
helpful discussions.
\end{acknowledgments}

\end{document}